\documentclass[%
 reprint,
showpacs,
 amsmath,amssymb,
 aps,
]{revtex4-1}

\usepackage{graphicx}% Include figure files
\usepackage{dcolumn}% Align table columns on decimal point
\usepackage{subcaption}
\usepackage{epstopdf}

\usepackage{bm}% bold math
\begin{document}

\preprint{APS/123-QED}

\title{Using liner surface modes in ducts to make obstacles reflectionless}
\author{Maaz Farooqui}
%\email{author.one@someplace.edu}
\author{Yves Aur\'egan}
%\email{author.two@someplace.edu}
\author{Vincent Pagneux}
%\email{author.three@someplace.edu}
\affiliation{Laboratoire d'Acoustique de l'Universit\'e du Mans, Centre National de la Recherche Scientifique (CNRS), Le Mans Universit\'e, Avenue Olivier Messiaen, 72085 Le Mans Cedex 9, France}

\date{\today}% It is always \today, today,

\begin{abstract}
Acoustic cloaking for the suppression of backscattering inside ducts is proposed in the audible range where plane waves are curved around the object using the surface modes of the liner. 
It is numerically shown that a slowly varying resonant liner (e. g. based on an array of tubes) creates a zone of silence in which an object of arbitrary shape can be acoustically hidden for a wide range of frequencies.
And then, a resonant liner has deflecting properties without reflection of the wavefront, which are close to an ideal invisibility cloak. 
This kind of cloaking is effective in a wide frequency band and the cloaking band is a function of the impedance and height of the obstacle relative to the conduit.  
For smooth shaped obstacles, there is an ability of the object to help hide itself, which increases the cloaking frequency band (self-cloaking).
Dispersion effects lead to slow sounds and distortion of the wave phase.
\end{abstract}

\maketitle

%\tableofcontents

\section{Introduction}

Acoustic liners are widely used in ducts and they act according to two main principles: absorption, related to the real part of its impedance, and scattering of the acoustical waves by the changes of acoustical impedance, that can occur even with non-dissipative liner \cite{auregan2015}. As a result, playing with the reactive part of the liner can lead to unusual properties such as making an obstacle in a duct acoustically undetectable.

Invisibility cloaks has been subject of immense interest for rendering objects invisible to waves. 
 Pendry \cite{pendry2006controlling} demonstrated an invisible cloak that can hide arbitrary objects from electro-magnetic illumination. A similar method is further extended which led to the development of several acoustic cloaks in form of ground \cite{zigoneanu2014three}, carpet cloaks \cite{yang2016metasurface} and several others \cite{cummer2008scattering,cummer2007one,popa2011experimental,li2017arbitrary}. 
In realization of the acoustic cloaks, acoustic metamaterials \cite{liu2000locally} can be employed but they are efficient generally in a very narrow frequency band. Furthermore, they are very sensitive to dissipation. 

Recently, Gradient index metamaterial (GIM) demonstrated one possible bridge linking propagating waves and surface waves
\cite{sun2012gradient,xu2013broadband}. It was realised that GIMs inside a waveguide can be used for mode conversion,
converting the waveguide mode gradually into Surface wave mode without any scattering, or vice versa. Later, a one-dimensional cloak with gradient index metamaterials operating on microwave frequencies was proposed based on similar principle of mode conversion \cite{gu2015broadband}. Following a similar trend, spoof surface plasmons \cite{yang2015broadband,pors2012localized,kats2011spoof} were exploited for converting transmission modes to surface plasmon modes. One relevant application of this conversion to surface modes is broadband surface-wave transformation cloak \cite{xu2015broadband}.

In this work, we describe cloaks in duct capable of broadband acoustic waves guidance around arbitrary shaped obstacles. 
 It is shown that an axial variation of the impedance of the wall creates a zone of silence in a duct. 
 Any obstacle in this zone has no effect on wave propagation and becomes acoustically undetectable over a wide frequency range. 
 It is a non-conventional but efficient way of bending of waves with purely reactive impedance that transform progressively plane waves in the duct into surface waves over the liner.  This result in a bending of the acoustic wave towards the liner with a slower sound speed (Fig. \ref{fig:1}). Cloaking of several different obstacle shapes is demonstrated in this work. In all the results presented, the acoustic damping provided by the liner is neglected. Furthermore, it is shown in this work that a massive smoothly varying obstacle with shape similar to liner is capable of self-cloaking. This implies, that the external cloaking due to liner coupled with the self-cloaking phenomenon serves to improve the overall broadband performance of the cloaking system.

\section{\label{sec:2}Cloaking in ducts using the wall admittance}

\begin{figure}[!htb]
\centering  
  \includegraphics[width=0.8\columnwidth]{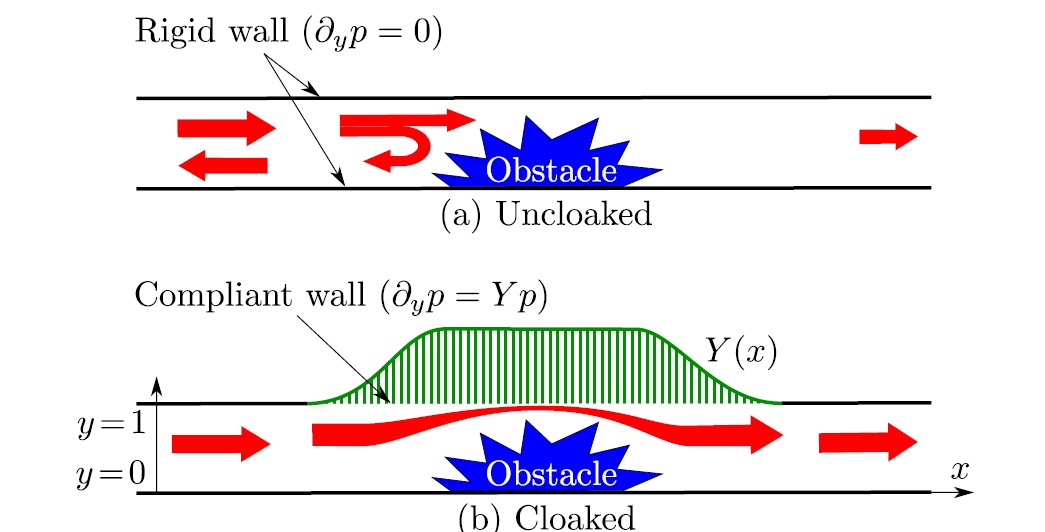}
  \caption{Schematic description of an obstacle in a duct with admittance  (a) 
  Uncloaked means a hard duct containing an arbitrary obstacle, (b) Cloaked means
  the same duct supplemented by a liner (compliant wall) 
  %a lined duct containing an arbitrary obstacle 
   }\label{fig:1}
                      \end{figure}

\begin{figure}[!htb]
  \centering  
   \includegraphics[width=0.8\columnwidth]{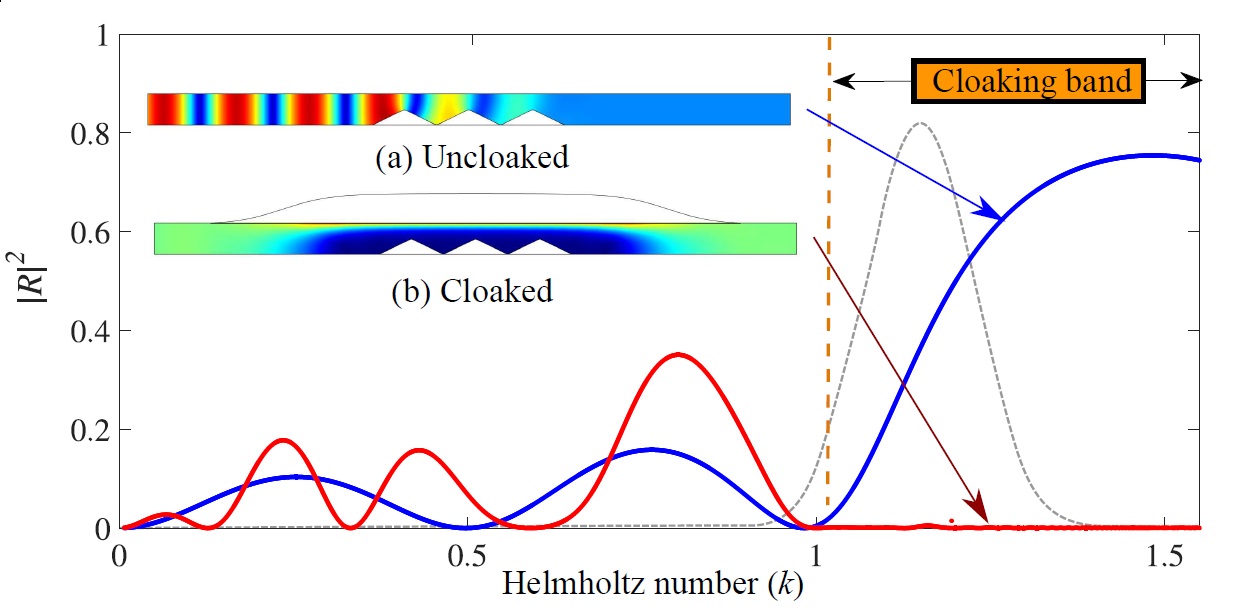}
   \caption{Reflection coefficient in energy $|R^2$| for  an array of triangular obstacles in the uncloaked (blue) and cloaked (red) cases. The grey dotted line is the Fourier transform in arbitrary unit of the pulse used in section \ref{Time_Domain}. (a) Absolute pressure field for the uncloaked case at $k$=1.38, (b) Absolute pressure for the cloaked case at $k$=1.38. }\label{fig:2}
   \end{figure}

We consider the sound propagation in a two dimensional channel, see Fig. \ref{fig:1}. The lower wall is rigid while the upper wall is compliant and described by a varying admittance $Y(x)$. When the distances are non-dimensioned by the height of the channel $H$, the Helmholtz equation, governing the propagation of the acoustic pressure $p$, is:
\begin{equation}
\label{eq:1}
\Delta p + {k^2} p = 0
\end{equation}
where $k =\omega H /c_0$ is the reduced frequency, $\omega$ is the frequency and $c_0$ is the sound velocity. The boundary conditions are $\partial_yp=0$ for $y=0$ and $\partial_y p=Y p$ for $y=1$ (Fig. \ref{fig:1}). For a uniform admittance $Y$, the solution is searched under the form $p = A \cosh(\alpha y) \exp(\mathrm{i}(-\omega t + \beta x))$ where $\alpha^2 = \beta^2-k^2$ and this leads to the dispersion relation:
\begin{equation}
\label{eq:2}
Y = {\alpha}\: \text{tanh}({\alpha})
\end{equation}
For a rigid wall ($Y=0$) at low frequencies, only the plane wave can propagate ($\alpha=0$). If the admittance is positive and, is slowly varying compared to the sound wavelength, 
the local value of
$\alpha$ increases progressively as $Y$ increases. It means that the wave is more and more concentrated against the wall, see Fig. \ref{fig:2}(b).  
The computations have been made by using COMSOL where only plane waves are incident from the left side of the duct. A silent zone is then created near the wall opposite to the admittance. Any obstacle located in this silent zone will have no influence on wave propagation. This obstacle is invisible in the sense that it does not produce any wave reflection. In order to verify the cloaking efficiency with a sharp scatterer, Fig. \ref{fig:2} depicts the reflection coefficient in energy $|R|^2$ for a series of triangular obstacles. The cloaking band in this case is from $1 \leqslant k \leqslant$ $\pi/2$.  On this example, the reflection coefficient of the 
uncloaked
obstacle is $|R|$ = 0.8549 while when the admittance is present it is reduced to $|R|$ = 0.02 in the cloaking band. In the sequel, the cloaking zone is defined as the frequency band where $|R|\leqslant 0.1 $.
 
\subsection{Frequency dependence}

\noindent Among others, a  realization of the admittance can be done by using small closed tubes 
perpendicular to the upper wall and with variable lengths $b(x)$. 
Considering lossless tubes, the admittance can be written as:
\begin{equation}
\label{eq:3}
Y (x,{k}) = k\: \tan(k \: b(x))
\end{equation}

When $k b\ll1$, the admittance can be approximated by $Y= k^2 b$ and the phase velocity of the wave is given by $c_\phi=(1+b)^{-1/2}$ meaning that the wave velocity is reduced compared to sound velocity \cite{auregan2015}. When the frequency $k$ goes to the first resonant frequency of the tubes given by $k_r b = \pi/2$, the admittance and $\alpha$ go to $\infty$ as $\alpha \simeq Y\simeq \pi/(2b(\pi/2-k b))$. Thus, for frequencies slightly below $k_r$, the wave decreases exponentially from the wall and has been transformed into a surface wave. Near $k_r$, the phase velocity of the wave goes linearly to 0 as $c_\phi \simeq (2 b/\pi )^2 (\pi/2-k b)$. The propagation is then highly dispersive.\\

The upper frequency of the cloaking band is given by the tube resonance ($k b = \pi/2$) while the lower frequency limit is a function of the object dimensions as well as the admittance. If change in the liner impedance is smooth enough to avoid liner reflection, the pressure must be small at the maximal height of the obstacle $h_0$ to have only a small reflection. It means that $ (1-h_0) \alpha \gg 1$ where $\alpha$ is given as a function of the admittance by Eq. (\ref{eq:2}).  

\section{Results for an obstacle in the shadow zone}

\subsection{Frequency Domain}

 \begin{figure}[!ht]
   \centering  
   \includegraphics[width=0.8\columnwidth]{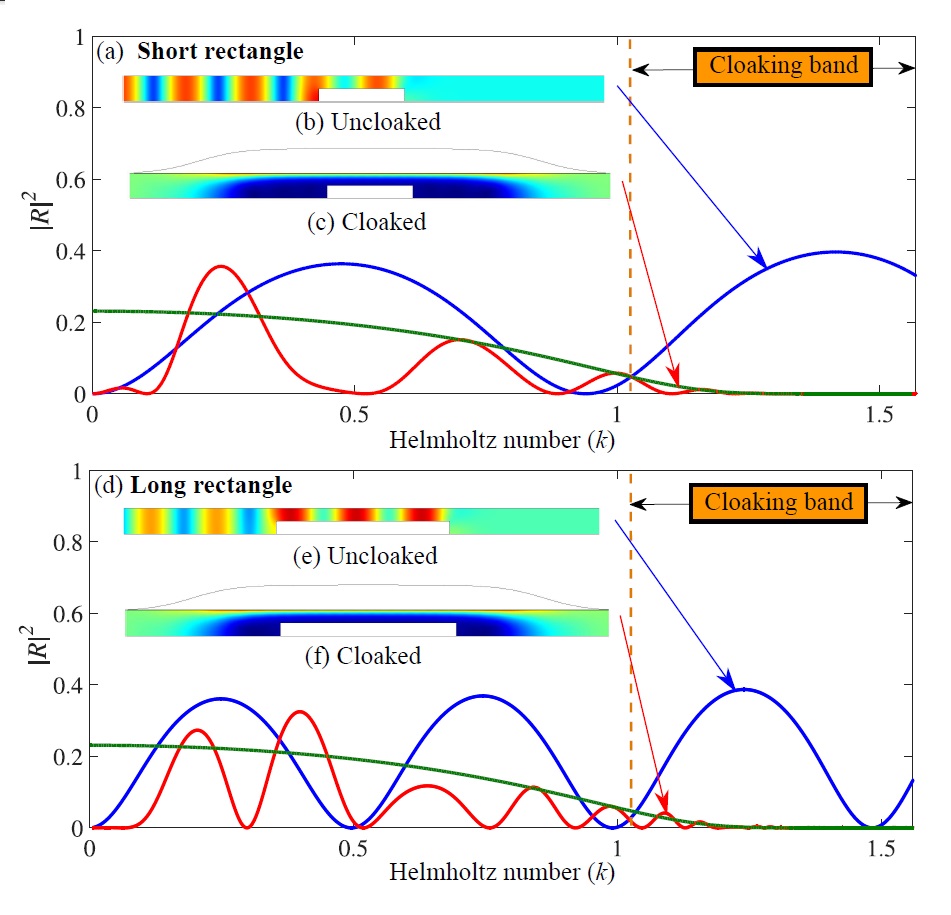}
     \caption{Comparison of the cloaking for short and long rectangular obstacles: (a, b $\&$ c) Short obstacle (d, e $\&$ f) Long obstacle. (a) and (d): $|R^2$| for uncloaked (blue) and cloaked (red) cases ($l=6$, $d= 2$, $b_0=1$).  (b, c, e $\&$ f) Absolute pressure field for $k=1.38$. The green curves in (a) and (d) is the value $|2R_\infty|^2$ that forms the envelope for case with uniform admittance treatment on an infinite rectangular obstacle}\label{fig:3}
   \end{figure}
   
A shadow zone can be created by taking an admittance profile under the form 
\begin{equation}
\label{eq:4}
b(x) = \frac{b_0}{2}\,\,\left( {\tanh \left( {\frac{{x + l}}{d}} \right) - \tanh \left( {\frac{{x - l}}{d}} \right)} \right)
\end{equation}
where the maximum tube height has been arbitrarily fixed to $b_0=1$ and the two parameters $l$ and $d$ allow to settle separately the maximal slope of the admittance change and the length of the shadow zone. In the following, those parameters are fixed to $l=6$ and $d= 2$ which correspond to an admittance variation smooth enough to avoid reflection for frequencies larger than $k=\pi/4$. 

Two rectangular obstacles with the same height ($h_0=0.5$) but different lengths ($w$=3 and 6) are put in the shadow zone, see Fig. \ref{fig:3}. For the short obstacle, (see Fig. \ref{fig:3}), the reflection coefficient at $k=1.38$ is reduced from $|R|$ = 0.6274 to $|R|$ = 0.0018 by application of the admittance. The reflection coefficient is reduced by the same amount for the long obstacle.  
Large oscillations of the reflection coefficient are seen without admittance. They are related to the interference of waves coming from the ascending step and waves coming from the descending step. 
Those two waves have the same amplitude and then $|R|$ is maximized by two times the reflection on the ascending step. 
When the change in admittance is smooth and when the admittance is almost constant over the rectangle, this procedure can also be applied to an obstacle with admittance. 
The absolute value of the reflection coefficient ($|R_\infty|$) of an ascending step in presence of a uniform admittance on the upper wall can be easily computed by a multimodal method. 
The value of $2 |R_\infty|$ is plotted in green on Fig. \ref{fig:3}(a) and (d). 
At low frequencies, there are some oscillations of the cloaked reflection coefficient (in red) over $2 |R_\infty|$ which are coming from some additional reflection on the admittance changes. 
But near the cloaking band, $2 |R_\infty|$ is a good envelope of the oscillations of $|R|$ when the admittance is present and can be used to evaluate with simple calculation the effect of the admittance.
 
This example shows that any object located in the shadow zone produces a small reflection on a broad frequency band even if this object covers the substantial surface of the shadow zone. An envelope of $|R|$ that can be computed from the maximum height of the object can be helpful to have a first idea of the acoustical behavior near and in the cloaking band.

\subsection{Time Domain}
\label{Time_Domain}

 \begin{figure}
     \centering
              \begin{subfigure}[b]{.45\textwidth}
	  \includegraphics[width=\columnwidth]{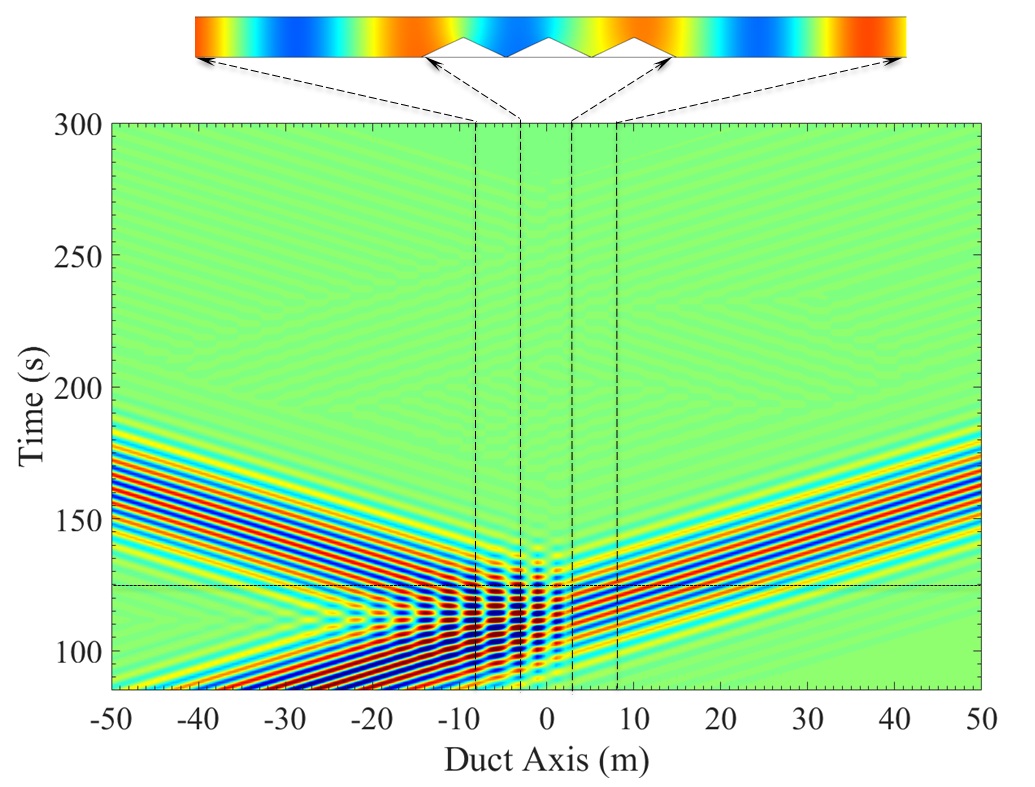}
         \caption{Triangular Obstactle uncloaked}\label{subfig-5:u}
       \end{subfigure}\\[\baselineskip]
           \bigskip
      \bigskip
    \begin{subfigure}[b]{.45\textwidth}
	  \includegraphics[width=\columnwidth]{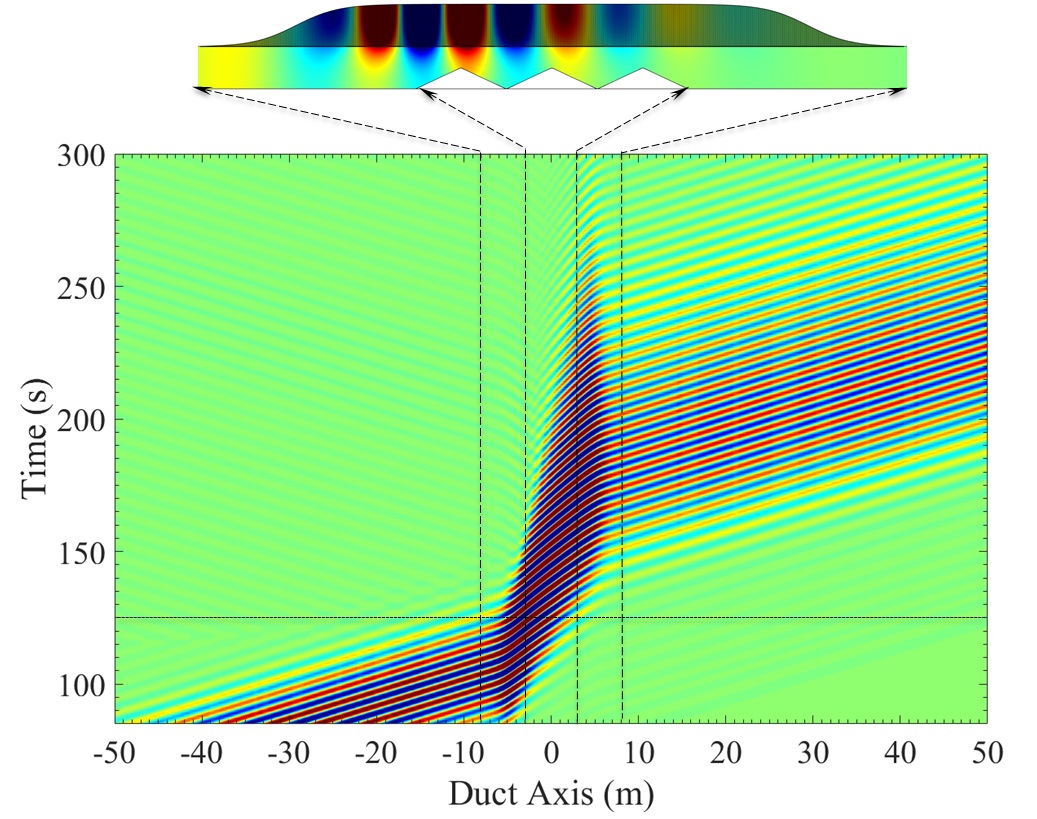}
       \caption{Triangular Obstactle Cloaked}\label{subfig-5:c}
     \end{subfigure}
                       \caption{Time-domain results for the pressure averaged over the channel with Gaussian-shaped wave packet on the left side for a duct with triangular obstacle (a) Uncloaked, (b) Cloaked.  Dotted line represents the time step (t=125 s) for pressure field depicted in each case, Dashed arrow correlates the dimensions of the plot with actual geometry}\label{fig:5}
   \end{figure}

To 
%better understand 
illustrate the broadband capability of
this cloaking effect, a time simulation of the configuration already shown in the frequency domain on Fig. \ref{fig:2} is performed using COMSOL. The obstacle of total length $w=6$ is made of 3 triangles of height $h_0=0.5$. This particular shape of the obstacle is chosen to be an efficient Bragg scatterer for frequencies such as $kd=n\pi$, where $n$ is an integer and $d=2$ is the distance between two triangles.  Accordingly, in the uncloaked case depicted in Fig. \ref{fig:2}, the reflection is high in the cloaking band (slightly below $k =\pi/2$). The admittance shape is the one used in the previous section (Eq.\ref{eq:4} with $b_0=1$, $l=6$ and $d= 2$).

A Guaussian wave packet ($\exp(-\frac{1}{2}(\frac{t-t_0}{s_t})^2)\sin(k_c t)$) is incident from the left side of the duct. The Fourier transform of this wave packet with a central frequency of $k_c= $ { 1.15} and with $s_s=$ 13 is plotted in Fig.\ref{fig:2} and is within the cloaked band. 

 The space-time variation of the pressure averaged over the channel is plotted in Fig. \ref{subfig-5:u} $\&$ \ref{subfig-5:c}, for uncloaked and cloaked cases, respectively. 
 The horizontal dotted lines on the plot corresponds to $t=125\,\,s$, for which the pressure field is depicted above the respective space-time plots. The vertical dashed lines on the plot corresponds to two zones, $(-3 \,\,{\leq} \,\,x \,\,{\leq} \,\,3)$ and $(-8.25 \,\,{\leq} \,\,x  \,\,{\leq} \,\,8.25)$. These two zones project the obstacle and the liner dimensions on the duct axis. 
 
 The wave reflection is very clearly visible in the uncloaked case and it has disappeared in the cloaked case. 
 In the cloaked case, the gradually slowing down of sound when entering in the cloaked region is visible. The slope of the contours shows that in the cloaking region, the sound propagation rate is reduced by a factor of 2.5 on average, while it propagates with a normalized sound velocity of $c_0=1$ elsewhere. And so, by passing through the cloaking region, the sound will be subject to a time delay compared to the uncloaked case. 
 The dispersion of the waves (resulting from the change in propagation speed as a function of frequency) can be visualized from (Fig. \ref{subfig-5:c}) by the widening of the transmitted wave packet. This effect is particularly pronounced in the presented case because, in order to have a low temporal width for a nice vizualisation, we have widened the frequency band that approaches the resonance where the variation of the wave velocity as a function of frequency is very large.

\section{Self-cloaking}

\begin{figure}[!htb]
 \centering  
   \includegraphics[width=0.8\columnwidth]{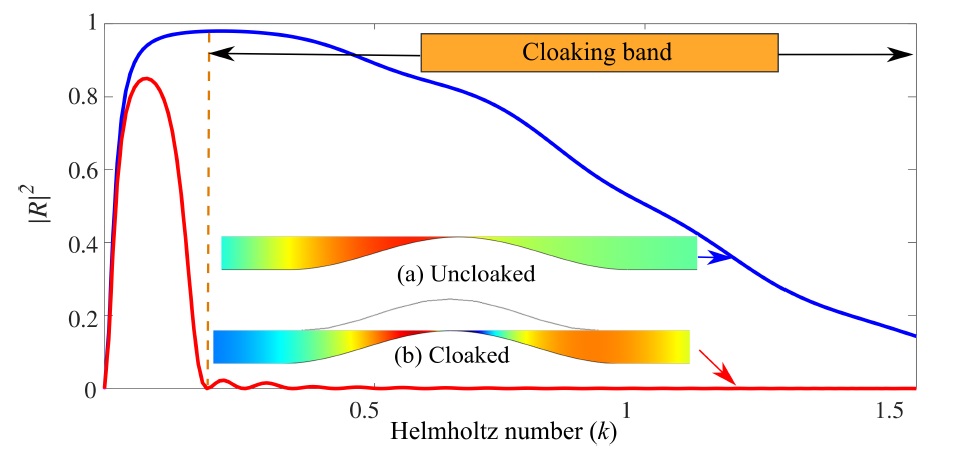}
   \caption{$|R^2|$ of a smooth cosine obstacle ($h_0=0.98$, $b_0=1$, $L=8.25$) for uncloaked (blue) and self-cloaked (red) cases. (a) Absolute pressure field for uncloaked case at $k$=0.25, (b) Absolute pressure for self-cloaked case at $k$=0.25. \label{fig:5}}
   \end{figure}

It has been seen in section \ref{sec:2} that the deflection of the wave towards the compliant wall is controlled by the parameter $\alpha$ which gives the exponential decrease in pressure from the wall. This parameter is directly related to $b$ which is the ratio of tube height to channel height. When an obstacle is present in the duct, the height of the channel between the obstacle and the wall decreases, $\alpha$ increases and consequently the presence of the obstacle helps to concentrate the wave towards the compliant wall. This effect is called self-cloaking and it improves the efficiency of the cloaking for smoothly varying obstacle. 

To be  demonstrative, even if the configuration is not very realistic due to the lack of losses consideration, an almost complete obstruction of the channel is considered. The shape of the obstacle is then chosen to be: 
\begin{equation}
\label{eq:3000}
h(x) = \frac{h_0}{2}(1+\cos(\frac{\pi{x}}{L}))
\end{equation}
where 
$h_0=0.98$ and 
$L=8.25$ and the height of the wall tubes is taken equal to the height of the obstacle. 
The results in the frequency domain for this geometry are given in Fig. \ref{fig:5}. Due to the particular geometry chosen, an almost complete reflection is obtained in the uncloaked case at low frequencies ($k\simeq0.25$). When an admittance is present, the reflection decreases drastically for the entire frequency band $0.25<k<\pi/2$. The cloaking band is then considerably increased compared to sharp obstacles where this self-cloaking effect cannot be seen due to the sudden change in the wave that induces reflection.

%\begin{figure}[!htb]
% \centering  
%   \includegraphics[width=0.8\columnwidth]{Fig_6.eps}
%   \caption{$|R^2|$ of a smooth cosine obstacle ($h_0=0.98$, $b_0=1$, $L=8.25$) for uncloaked (blue) and self-cloaked case with different cloaking lengths ($L_c$). $L_c=L/2$ (green), $L_c=L$ (red), $L_c=3L/2$ (black) and $L_c=2L$ (brown). \label{fig:6} } 
%\end{figure} 
%                     
\section{Conclusion}

In this paper, we construct a broadband acoustic cloak to reduce backscattering inside ducts using liner surface modes. 
By using a smooth variation in wall admittance of a duct (which does not produce reflection), it is possible to create a silent zone in which acoustic waves do not penetrate. 
When an object is positioned in this silent zone, it has no influence on the wave propagation and this object is undetectable by acoustic waves. 
This means that the amplitude of the reflection coefficient of the object in the silent zone is equal to 0 and the amplitude of the transmission coefficient is equal to 1.  
This cloaking effect always exists for frequencies slightly lower than the liner resonance and the low limit frequency of this effect is related to the height of the obstacle. 
For an object that is half the height of the duct, this effect occurs at frequencies between $f_R$, the resonance frequency of the liner and $2 f_R/3$, which is a very wide frequency band compared to other conventional cloaking techniques.
 When the object is smooth enough, the presence of the object helps the cloaking to occurs on a wider frequency bandwidth. This effect is called self-cloaking.

The simplicity of this cloaking, made simply from a resonant wall, contrasts with other cloaking techniques that are more complex to perform and that only work at a single frequency.
The creation of a zone of silence in a duct and the concentration of the wave along a wall can also have other applications than cloaking, such as the realization of an anechoic termination \cite{Maaz2018}.

%\lipsum[1]
\section*{Acknowledgements}

\noindent This work was supported by the International ANR project FlowMatAc a co-operation project between France and Hong Kong.
%\section*{References}

%\bibliographystyle{elsarticle-num}
%\biboptions{numbers,sort&compress}
\bibliography{archive_JSV}

\end{document}